\documentclass{article}
\usepackage{spconf,amsmath,epsfig}
\usepackage{algorithm}
\usepackage{algorithmic}
\usepackage{graphicx, subfigure}

\title{EAST: Energy-efficient Adaptive Scheme for \\Transmission in Wireless Sensor Networks}

\name{M. Tahir$^{\ddag}$, N. Javaid$^{\ddag}$, Z. A. Khan$^{\$}$, U. Qasim$^{\sharp}$, M. Ishfaq$^{\S}$}

\address{$^{\ddag}$COMSATS Institute of Information Technology, Islamabad, Pakistan. \\
        $^{\$}$Faculty of Engineering, Dalhousie University, Halifax, Canada.\\
        $^{\sharp}$University of Alberta, Alberta, Canada.\\
        $^{\S}$King Abdulaziz University, Rabigh, Saudi Arabia.}

\begin{document}
%
\maketitle
\begin{abstract}
In this paper, we propose Energy-efficient Adaptive Scheme for Transmission (EAST) in WSNs. EAST is IEEE 802.15.4 standard compliant. In this approach, open-loop is used for temperature-aware link quality estimation and compensation. Whereas, closed-loop feedback helps to divide network into three logical regions to minimize overhead of control packets on basis of Threshold transmitter power loss $(RSSI_{loss})$ for each region and current number of neighbor nodes that help to adapt transmit power according to link quality changes due to temperature variation. Simulation results show that propose scheme; EAST effectively adapts transmission power to changing link quality with less control packets overhead and energy consumption compared to classical approach with single region in which maximum transmitter power assigned to compensate temperature variation.
\end{abstract}

\section{Related Work and Motivation}
\label{sec:format}

To transmit data efficiently over wireless channels in WSNs, existing schemes set some minimum transmission power for maintaining reliability. These schemes either decrease interference among nodes or unnecessary energy consumption. In order to adjust transmission power, reference node periodically broadcasts a beacon message. When neighbor nodes hear a beacon message from a reference node, neighbor nodes transmit an ACK message. Through this interaction, reference node estimate connectivity between neighbor nodes \cite{1}.

In Local Mean Algorithm (LMA), a reference node broadcasts LifeMsg message. Neighbor nodes transmit LifeAckMsg after they receive LifeMsg. Reference nodes count number of LifeAckMsgs and transmission power is controlled by maintaining appropriate connectivity. For example if number of LifeAckMsgs is less than NodeMinThresh transmission power is increased. In contrast, if number of LifeAckMsgs is more than NodeMaxThresh transmission power is decreased. As a result, they provide improvement of network lifetime in a sufficiently connected network. However, LMA only guarantees connectivity between nodes and cannot estimate link quality \cite{2}.

Local Information No Topology/Local Information Link-state Topology (LINT/LILT) and Dynamic Transmission Power Control (DTPC) uses $RSSI_{loss}$ to estimate transmitter power level. Nodes exceeding Threshold $RSSI_{loss}$  are regarded as neighbor nodes with reliable links. Transmission power also controlled by Packet Reception Ratio (PRR) metric \cite{3}.

Since $RSSI_{loss}$ is directly proportional to temperature. Adaptive Transmission Power Control (ATPC) adjusts transmission power dynamically according to spatial and temporal effects. This scheme tries to adapt link quality that changes over time by using closed-loop feedback. However, in large-scale WSNs it is difficult to support scalability due to serious overhead required to adjust transmission power of each link ~\cite{4}. Existing approaches estimate variety of link quality indicators by periodically broadcasting a beacon message. In addition, feedback process is repeated for adaptively controlling transmission power. In adapting link quality for environments where temperature variation occur, packet overhead for transmission power control should be minimized. Reducing number of control packets while maintaining reliability is an important technical issue \cite{5}.

\section{Proposed Energy Efficient Adaptive Transmission Scheme}
\label{sec:pagestyle}

Our transmission power control scheme is designed to efficiently combine closed-loop and open-loop feedback processes to divide network into three logical regions. It utilizes open-loop process based on sensed temperature information to reduce overhead for transmission power control according to temperature variation. Closed-loop feedback process based on control packets is further used to accurately adjust transmission power. By adopting both open-loop and closed-loop feedback processes we divide network into three regions, A for High $RSSI_{loss}$, B for Medium $RSSI_{loss}$ and  C for Low $RSSI_{loss}$  on basis of Threshold $RSSI_{loss}$  for each region.

In order to assign minimum and reachable transmission power to each link EAST is designed. EAST has two phases that is initial and run-time. In initial phase, each node build a model for each of its neighbors links. The run-time phase based on previous model EAST adapts the link quality to dynamically maintain each link with respect to time. In a relatively stable network, control overhead occurs only in measuring link quality in initial phase. In a relatively unstable network because link quality is continuously changing initial phase is repeated and serious overhead occur. Before we present block diagram for our propose scheme some variables are defined as follows: (1) Number of current neighbor nodes $n_{c}(t)$, (2) Desired number of neighbor nodes $n_{d}(t)$, (3) Error: e(t) = $n_{d}(t) - n_{c}(t)$,(4) Transmission power level $P_{level}$.

In order to adjust transmission power, transmission power level determined as connectivity with neighbor nodes. After comparing number of current neighbor nodes with a set point desired number of neighbor nodes power controller adjusts transmission power level accordingly. PRR, ACK, and $RSSI_{loss}$ used to determine connectivity. ACK estimate connectivity, it cannot determine link quality. PRR estimate connectivity accurately that causes significant overhead \cite{6}. In our scheme, we use $RSSI_{loss}$ for connectivity estimation, which measure connectivity with relatively low overhead.

Power controller adjust transmission power level by utilizing both number of current neighbor nodes and temperature sensed at each neighbor node. Since power controller is operated not merely by comparing number of current neighbor nodes with desired number also by using temperature-compensated power level, so that it reach to desired power level rapidly. If temperature is changing then temperature compensation is executed on basis of relationship between temperature and $RSSI_{loss}$. Network connectivity maintained with low overhead by reducing feedback process between nodes which is achieved due to logical division of network while link quality is changing due to temperature variation.

Transmission power loss due to temperature variation formulated using relationship between $RSSI_{loss}$ and temperature experimented in Bannister et al.. Mathematical expression for $RSSI_{loss}$ due to temperature variation is as follows \cite{7}:

\begin{equation}
RSSI_{loss}[dBm]=0.1996*(T[C^{o}]-25[C^{o}])
\end{equation}

To compensate $RSSI_{loss}$ estimated from Eq.(1) we have to control output power of radio transmitter accordingly. Relationship between required transmitter power level and $RSSI_{loss}$ is formulated by Eq.(2) using least square approximation \cite{7}:

\begin{equation}
P_{level}[dBm]=[(RSSI_{loss}+40)/12]^{2.91}
\end{equation}

Based on Eqs (1, 2), we obtain appropriate power level to compensate $RSSI_{loss}$ due to temperature variation. To compensate path loss due to distance  between each sensor node in WSN free space model help to estimate actual required transmitter power. After addition of required power level due to temperature variation and distance given in Eq.(3), we estimate actual required transmitter power between each sensor node. For free space path loss model  we need number of nodes, distance between each node, required $E_{b}/N_{o}$ depends upon $SNR$, spectral efficiency $\eta$, frequency $f$ and receiver noise figure $(RNF)$ required:

 \begin{equation}
P_{t}[dBm]=[\eta*(E_{b}/N_{0})*mkTB*(4\pi d /\lambda )^2+RNF]+P_{level}
\end{equation}

 Our scheme aims to simplify transmission power control by compensating $RSSI_{loss}$ change based on temperature information sensed at each node. Propose compensation scheme does not require any communication overhead with neighbor nodes, rather utilizes information gathered from temperature sensor. Open-loop control reduce significantly complexity of closed-loop feedback control for transmission power control. We define important parameters for our propose scheme,(1) $RSSI_{loss}$ Threshold for each region. (2) Number of desired neighbor nodes in each region $n_{d}(t)=n_{c}(t)-5$, (3) Transmission power level for each region.

Threshold $RSSI_{loss}$ is minimum value required to maintain link reliability. Reference node broadcasts beacon message periodically to neighbor nodes and wait ACKs. If ACKs are received from neighbor nodes then $RSSI_{loss}$ is estimated for logical division of network, number of nodes with high $RSSI_{loss}$  considered in region A, medium $RSSI_{loss}$ considered in region B, and with low $RSSI_{loss}$ in region C. If ($RSSI_{loss}$ $\geq$ $RSSI_{loss}$ Threshold) and  ($N_{current}\geq  N_{desired}$) then Threshold transmitter power level assigned if  for similar case ($N_{current}<  N_{desired}$) then similar transmitter power assigned and if  ($RSSI_{loss}$ $<$ $RSSI_{loss}$  Threshold) then by default keep same transmitter power level. Given below is complete algorithm for EAST.

\section{Results and Discussion}
\label{sec:typestyle}

In this section we describe simulation results of our propose technique for power efficient transmission in WSNs compared with classical approach used for transmission of data. In Fig1 we have shown values of meteorological temperature for one round that each sensor node have sensed in WSN. Let suppose we have 100 nodes in 100*100 $m^{2}$ square region and temperature have values in range (-10 - 53)$C^{o}$  \cite{8} for given meteorological condition of Pakistan. Reference node is placed at edge of this region. In figure shown earlier temperature variation shown on y-axis and corresponding nodes on x-axis. Each sensor node placed at different location randomly in given area and we clearly see variation of temperature for different nodes in WSN.

 \begin{figure}[h]
\begin{center}
\includegraphics[scale=0.25]{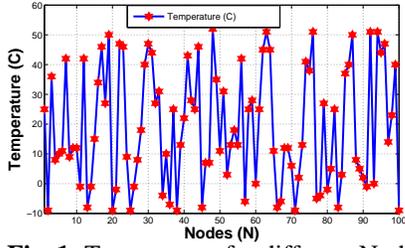}
\vspace{-0.5cm}
\caption{Temperature for different Nodes}
\end{center}
\end{figure}

 \begin{figure}[h]
\begin{center}
\includegraphics[scale=0.25]{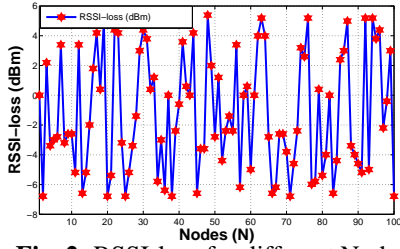}
\vspace{-0.5cm}
\caption{RSSI-loss for different Nodes}
\end{center}
\end{figure}

 \begin{figure}[h]
\begin{center}
\includegraphics[scale=0.25]{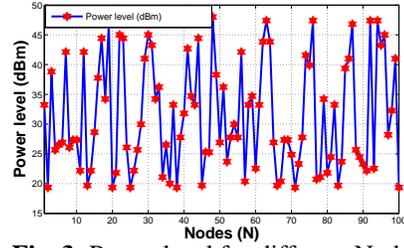}
\vspace{-0.5cm}
\caption{Power level for different Nodes}
\end{center}
\end{figure}

 \begin{figure}[h]
\begin{center}
\includegraphics[scale=0.25]{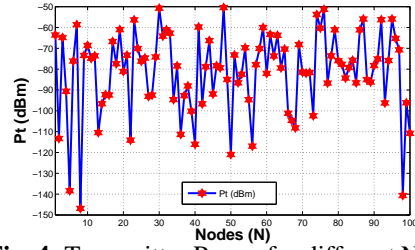}
\vspace{-0.5cm}
\caption{Transmitter Power for different Nodes}
\end{center}
\end{figure}

Different values of temperature for each sensor node based on meteorological condition  help to estimate $RSSI_{loss}(dBm)$ that is transmitter power loss. Fig2 shows transmission power loss due to temperature variation in any environment using the relationship between $RSSI_{loss}(dBm)$ and temperature $(C^{o})$ given by Bannister et al. $RSSI_{loss}(dBm)$  on y-axis indicates transmission power loss for each  sensor node. $RSSI_{loss}(dBm)$ high means that sensor node  placed in region where temperature is high so link not have good quality. For temperature (-10 - 53)$C^{o}$  $RSSI_{loss}(dBm)$ have value in range (-6dBm) - (5dBm). Chosen simulation parameters are Rounds 1200, Temperature -10-53 $C^{0}$, Distance (1-100)m, Nodes 100, Regions A,B,C, $\eta$ 0.0029, SNR 0.20dB, Bandwidth 83.5MHz, Frequency 2.45GHz, RNF 5dB, T 300k, $E_{b}/N_{0}$  8.3dB.

\begin{table}[h!]
  \centering
  \caption{Estimated Parameters}
  \tiny
  \begin{tabular}{|c|c|}\hline
    Number of Nodes (A,B,C)   & 46,30,24 \\ \hline
    Desired Neighbors  (A,B,C)   & 41,25,19 \\ \hline
     Nodes after 1200 Rounds (A,B,C) & 41,22,17 \\ \hline
    Threshold power level (A,B,C)  & 43.24,31.77,22.21 dBm \\ \hline
    Nodes above threshold $RSSI_{loss}$ (A,B,C)  & 23,11,8 \\ \hline
    Nodes below threshold $RSSI_{loss}$ (A,B,C)   & 18,11,9 \\ \hline
    PRR (A,B,C) & (80-98),(70-96),(63-97) $\%$ \\ \hline
    Threshold $RSSI_{loss}$ ( A,B,C) & 3.78,-0.61,-5.17 dBm \\\hline
    \end{tabular}%
  \label{tab:addlabel}%
\end{table}%

From figure shown earlier it is also clear that link quality and $RSSI_{loss}$ have inverse relation, when temperature is high $RSSI_{loss}$ has high value means low quality link and vise versa. As we have earlier mentioned link quality and $RSSI_{loss}$ have inverse relation that is for high temperature link quality is not good and for low temperature link quality is good. After estimating $RSSI_{loss}$ for each node in WSN we compute corresponding transmitter power level to compensate $RSSI_{loss}$. $P_{level}$ assigned to each node on basis of nodes estimated $RSSI_{loss}$. Fig2 shows range of power levels on y-axis for given $RSSI_{loss}$ that is between (20- 47) $dBm$ and also variation of required power level for sensor node with changing temperature that is at low temperature required $P_{level}$ is low and for high temperature required $P_{level}$ is high.

 \begin{figure}[h]
\begin{center}
\includegraphics[scale=0.25]{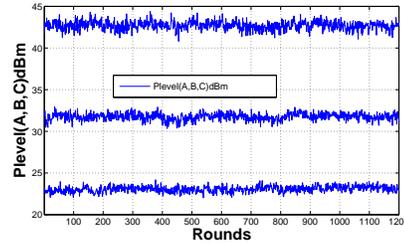}
\vspace{-0.5cm}
\caption{ Power level  for different regions}
\end{center}
\end{figure}

As we have earlier estimated $RSSI_{loss}$ for each sensor node on the basis of given meteorological temperature that help to estimate required power level to compensate transmission power loss. That power level only help to compensate $RSSI_{loss}$ due to temperature variation. To compensate path loss due to distance  between each sensor node in WSN free space model help to estimate actual required transmitter power. After addition of required power level due to temperature variation and distance, we estimate actual required transmitter power between each sensor node. Fig3 shows required transmitter power including both transmission power loss due to temperature variation and free space path loss for different nodes. We clearly see from figure that $P_{t}$ lies between (-115 - 45)$dBm$ and most of times it is above -100$dBm$ .

 \begin{figure}[h]
\begin{center}
\includegraphics[scale=0.25]{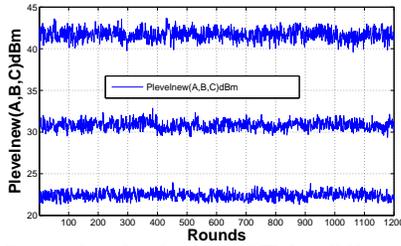}
\vspace{-0.5cm}
\caption{Power level using $EAST$ for different regions}
\end{center}
\end{figure}

As we have chosen 1200 rounds for our analysis each round starts when temperature change detected also we have divided network into three regions (A,B,C) for analysis of our propose technique. Table1 below shows estimated parameters like number of nodes in each region based on $RSSI_{loss}$, Threshold $RSSI_{loss}$ for each region, nodes above and below threshold in each region, packet reception ratio for each region based on current number of neighbor nodes and desired number of nodes and threshold power level for each region after 1200 rounds. From sensor nodes sensed temperature we have estimated $RSSI_{loss}$ that describes transmission power loss due to temperature variation. After that we have assigned $RSSI_{loss}$ to each node. In our approach we have divided network into three regions on basis of Threshold $RSSI_{loss}$ and count numbers of nodes in each region. Nodes with high $RSSI_{loss}$ in region (A), medium $RSSI_{loss}$ in (B) and low $RSSI_{loss}$ in (C).

After estimating $RSSI_{loss}$ for nodes of each region we have estimated required $P_{level}$ for nodes of each region that we clearly see in Fig5, in region A $P_{level}$ lies between (40-45)$dBm$, for region B (30-35) $dBm$ and for region C (20-25)$dBm$. It means that for region A required power level high then both other region that also shows that for that region temperature and $RSSI_{loss}$ is large. For region B required power level is between both region A and C and for C region required power level is less then both other two regions. We have earlier seen in Fig5 power level for each region assigned using classical approach. After applying our propose technique we see what power level required for each region. We clearly see difference  between $P_{level}$ as shown in Fig6, that required power level decrease for each region and for region A it decreases maximum.

\section{Conclusion}
\label{sec:majhead}

In this paper, we have presented  our propose technique  EAST to study temperature effect on wireless link quality. It shows that temperature is one of most important factors impacting link quality variation. Relationship between $RSSI_{loss}$ and temperature has been analyzed for our transmission power control scheme. Our scheme uses open-loop feedback control to compensate for changes of link quality according to temperature variation. By combining both open-loop temperature-aware compensation and close-loop feedback control, we significantly reduce overhead of transmission power control in a WSN. We  further extended our scheme by dividing network into three regions on basis of Threshold $RSSI_{loss}$ and assign power level to each node in three regions on basis of current number of nodes and desired number of nodes, which help to adapt transmitter power according to link quality variation and increase network lifetime.

\end{document}